\documentclass[11pt]{article}
\usepackage{amssymb}
\usepackage{euscript}

\def\a{{\alpha}}

\def\d{\partial}

\def\n{{\bf n}}
\def\BN{{\bf N}}
\def\BM{{\bf M}}

\def\m{{\bf m}}
\def\bra#1{\langle #1 |}
\def\ket#1{|#1 \rangle}
\def\0{\nonumber}

\def\det{{\rm det}}

\def\log{{\rm log}}
\def\tr{{\rm tr}}

\def\exp{{\rm exp}}

\newcommand\N{{\cal{N}}}

\newcommand\ee{\end{eqnarray}}      
\newcommand\be{\begin{eqnarray}}
\newcommand\ba{\begin{array}}           
\newcommand\ea{\end{array}}
\newcommand\eeq{\end{equation}}     
\newcommand\beq{\begin{equation}}

\begin{document}
\begin{flushright}
SISSA/54/06/EP\\ ULB--TH/06-18\\ hep-th/0609182
\end{flushright}

\vspace{.1in}
\begin{center}
{\LARGE\bf  Towards open--closed string duality}: \\
\vspace{5mm}
{\Large\bf Closed Strings as Open String Fields}
\end{center}
\vspace{0.1in}
\begin{center}
L. Bonora $^{(a)}$\footnote{ bonora@sissa.it}, N. Bouatta
$^{(b)}$\footnote{Chercheur FRIA, nbouatta@ulb.ac.be},  C.
Maccaferri $^{(c)}$\footnote{carlo.maccaferri@ulb.ac.be}
\vspace{7mm}

$^{(a)}$ {\it International School for Advanced Studies (SISSA/ISAS)\\
Via Beirut 2--4, 34014 Trieste, Italy, and INFN, Sezione di
Trieste} \vspace{5mm}

$^{(b)}$  {\it Physique Th\'eorique et Math\'ematique,
Universit\'e Libre de Bruxelles \& International Solvay
Institutes, ULB Campus Plaine C.P. 231, B--1050 Bruxelles,
Belgium}\vspace{5mm}

$^{(c)}$ {\it Theoretische Natuurkunde, Vrije Universiteit Brussel,
Physique Th\'eorique et Math\'ematique, Universit\'e Libre de Bruxelles,
and
The International Solvay Institutes
Pleinlaan 2, B-1050 Brussels, Belgium
}
\end{center}
\vspace{0.1in}
\begin{center}
{\bf Abstract}
\end{center}

We establish a translation dictionary between open and closed
strings, starting from open string field theory. Under this
correspondence, (off--shell) level--matched closed string states
are represented by star algebra projectors in open string field
theory. Particular attention is paid to the zero mode sector,
which is indispensable in order to generate closed string states
with momentum. As an outcome of our identification, we show that
boundary states, which in closed string theory represent
D--branes, correspond to the identity string field in the open
string side. It is to be remarked that closed string theory
D--branes are thus given by an infinite superposition of star
algebra projectors.


\vspace{0.2in} \noindent Keywords: String Field Theory,
Open--closed string duality \vspace{0.2in}
\newpage
\section{Introduction}

The duality between open and closed strings has been a well--known
topic since the very beginning of string theory. The AdS/CFT
correspondence has revived the interest on this subject: it is a
sort of limiting case in which the open string side of the
correspondence is represented by a conformal gauge theory. More
recently A. Sen, \cite{Sen}, has suggested that open string theory
might be able to describe all the closed string physics, at least
in a background where D--branes are present. In this sense
Witten's open string field theory should be a privileged ground to
check this idea. For open string field theory is of course
formulated in terms of open strings degrees of freedom, but there
is ample evidence that tachyon condensation leads to a new vacuum
and that this new vacuum is the closed string one.

In a recent paper \cite{BMST2} a remarkable correspondence was pointed
out (in the context of the AdS/CFT duality) between N=4 SYM states in 4D
and star algebra projectors, or, more appropriately,
family of them and star algebra projectors in SFT (which can be interpreted
also as VSFT solutions).
Upon taking a coarse graining limit, the former give rise to the geometry
of supergravity solutions (the 1/2 BPS solutions of
\cite{LLM}). Although the correspondence is imperfect due to the lack of
supersymmetry on the SFT side, it is very suggestive, because it implies
that supergravity solutions can be constructed out of open string bricks.
Logically one expects that closed string modes should be expressible in terms
of open string degrees of freedom. Following Sen's suggestion and this
example we set out in this paper to tackle the problem of writing out an
explicit relation between open and closed string modes, a sort of dictionary
to translate from the open string to the closed string language. Our proposal
can be summarized as in the title: (perturbative) closed string modes are
SFT projectors. More precisely: momentum and level--matched off--shell
closed string states are in one--to--one correspondence with
star algebra projectors in SFT. In this paper we start also to verify the
validity of this dictionary. One very interesting outcome
is the proof that (if we neglect localization, which is taken care of by the
zero mode sector) a boundary state describing a D--brane in the closed
string language under this correspondence gets translated into the open
string identity state. This implies in particular that such boundary states
are superpositions of infinite many star algebra projectors

Before we proceed to expound our paper it is necessary for us to
make a short digression to comment the state of affairs in SFT,
which is actually very fluid and promising. Recently M. Schnabl
\cite{Schn2,Schn3} has found an exact analytic solution to the SFT
equation of motion, which corresponds to a vacuum without
perturbative open strings modes, see also \cite{Oka,FK1,FK2}. More
solutions of this type have been found in \cite{RZ}. These prove
the first two Sen's conjectures \cite{Senconj}. For the third one,
existence of lower dimensional brane solutions, more work is
needed. The existence of such solutions was shown in the past in
the context of the VSFT \cite{Ras}, a simplified (and singular)
version of Witten's open SFT, which is likely to give the correct
response at least for static solutions and it may be that to any
such VSFT solution there correspond an analytic SFT solution \`a
la Schnabl. We recall that the solutions to the VSFT equation of
motion are star algebra projectors (at least for the matter part).
In this paper the basic objects are precisely star algebra
projectors. Since the star product is the same in SFT and in VSFT,
star algebra projectors are well defined objects in SFT, even
without reference to VSFT. This is the sense in which they will be
considered in this paper, namely as objects pertaining to SFT. In
this regard a warning is in order: generally speaking the
(perturbative) closed string states we introduce in this paper are
star projectors, that is D--branes in the VSFT interpretation,
while D--branes (or boundary states) in closed string theory
correspond to infinite superposition of the latter.

The paper is organized as follows. Section 2 and 3 are essentially
preliminary and contain mostly well--known results about split string
field theory and the comma vertex algebra, although elaborated the way
we need. In section 4 we establish the conjectured correspondence
between zero momentum closed string states and star algebra projectors.
In section 5 we associate to such states a momentum eigenfunction and
conclude that the resulting (off--shell) closed string states are
again star algebra projectors. In section 6 we show that a boundary state
describing a D--brane in the closed string language gets translated
into the open string identity state, and suggest an interpretation of
this fact. As a test of our construction we also show how to compute
the closed string exchange between two boundary states using standard star
algebra manipulations. Finally, section 7 is devoted to a discussion of our
results and of the questions they raise.

\section{Preliminary algebras}

The SFT action is
\beq
{\cal S}(\Psi)= -  \left(\frac 12 \langle\Psi
|{Q_B}|\Psi\rangle +
\frac 13 \langle\Psi |\Psi *\Psi\rangle\right)\label{sftaction}
\eeq
where $Q_B$ is the open string BRST charge.

We will consider star algebra projectors, that is states
that satisfy the equation
\be
\Psi * \Psi = \Psi\0
\ee
Since the star product factorizes into matter and ghost part
it is natural to make for projectors the following factorized ansatz
\beq
\Psi= \Psi_m \otimes \Psi_g\label{ans}
\eeq
where $\Psi_g$ and $\Psi_m$ depend purely on ghost and matter
degrees of freedom, respectively. The projector equation   splits into
\be
\Psi_g & = &  \Psi_g *_g \Psi_g\label{PEg}\\
\Psi_m & = & \Psi_m *_m \Psi_m\label{PEm}
\ee
where $*_g$ and $*_m$ refer to the star product involving only the ghost
and matter part.

We will concentrate on the matter part, eq.(\ref{PEm}).
The $*_m$ product in the operator formalism is defined as follows
\beq
_{123}\!\langle V_3|\Psi_1\rangle_1 |\Psi_2\rangle_2 =_3\!\langle
\Psi_1*_m\Psi_2|,
\label{starm}
\eeq
see \cite{GJ1,Ohta, tope, leclair1} for the definition of the three
string vertex $_{123}\!\langle V_3|$. The basic ingredient in this
definition
are the matrices of vertex coefficients $V_{nm}^{rs}$, $r,s=1,2,3,\,\,
n,m=1,\ldots,\infty$.

The following developments are based on the sliver solution.
\beq
|\Xi\rangle = \N e^{-\frac 12 a^\dagger\cdot S\cdot a^\dagger}|0\rangle,
\quad\quad
a^\dagger\cdot S \cdot a^\dagger = \sum_{n,m=1}^\infty a_n^{\mu\dagger}
S_{nm} a_m^{\nu\dagger}\eta_{\mu\nu}\label{Xi}
\eeq
where $S= CT$ and
\beq
T= \frac 1{2X} (1+X-\sqrt{(1+3X)(1-X)})\label{sliver}
\eeq
with $X=CV^{11}$, where $C_{nm}=(-1)^n\delta_{nm}$ is the twist
matrix.
The normalization constant $\N= \left(\sqrt{\det(1+T)(1-X)}\right)^D$
is formally vanishing and needs to be regularized.
It has been showed in other papers how this and related problems could be
dealt with, \cite{BMP1, BMP2}. Our basic projector will have the form of the
sliver along the the space--time directions.

In SFT there are two preferred ways to split the set of open string
oscillators in two distinct sets which could mimic
the two sets of holomorphic and anti--holomorphic closed string oscillators:
the even--odd and left--right splitting.
The former consists in splitting according to the eigenvalues of the twist
matrix $C$; one can quickly verify, however, that it does not fit our purposes.
The latter is based on the separation between the left and right modes of the
open string. This turns out to be a better chance.

The construction that follows is based on split string field theory and
the so--called comma vertex algebra,
\cite{Chan,Bordes,Abdu,RSZ3,GT,Moeller,Furu}. Many formulas
we use in this section can be traced back to ref. \cite{Furu}.

Let us introduce the left and right Fock space projector $\rho_L$ and
$\rho_R$:
\be
\rho_L ^2=\rho_L,\quad\quad \rho_R^2=\rho_R,\quad\quad \rho_L+\rho_R=1
\label{RL}
\ee
with the properties
\be
\rho_L= \rho_L^t = C\rho_R C,\quad\quad \rho_R= \rho_R^t = C\rho_L C
\ee
They turn out to project onto the left and right hand part of the string
respectively. Next we define the operators
\be
s^\mu = \omega (a^\mu +S a^{\mu\,\dagger})= (a^\mu +S a^{\mu\,\dagger})\omega,
\quad\quad\omega =\frac 1{\sqrt{1-S^2}}\label{Bog}
\ee
and the conjugate ones,
where the labels $n,m$ running from 1 to +$\infty$ are understood.
Using the algebra of open string creation and annihilation operators these
operators can be shown to satisfy
\be
[ s_m^\mu,s_n^{\nu^\dagger} ]= \delta_{nm} \eta^{\mu\nu}\label{sscommm}
\ee
while the other commutators vanish.
Moreover, understanding the Lorentz indexes,
\be
s_n |\Xi\rangle= \N e^{-\frac 12 a^\dagger S a^\dagger}\omega(a-Sa^\dagger
+Sa^\dagger)|0\rangle=0\label{Bogvac}
\ee
Therefore the combinations $s_n$ represent Bogoliubov transformations, which
map the Fock space based on the initial vacuum $|0\rangle$ to a new Fock
space in which the role of vacuum is played by the sliver.

Now we introduce the vector $\xi$ such that
\be
\rho_L \xi=\xi, \quad \quad \rho_R \xi=0\label{xi}
\ee
As a consequence
\be
\rho_R C\xi = C\xi, \quad\quad \rho_L C\xi=0\0
\ee
There exists a complete basis $\xi_n$ ($n=1,2,...$) that satisfy these
conditions and are orthonormal in the sense that
\be
\langle \xi_n|\frac 1{1-T^2} |\xi_m\rangle = \delta_{nm}\label{normal}
\ee
see, for instance, \cite{BMP1}.

Let us define, for any $\xi$,
\be
\xi^{L} = \frac 1{\sqrt{1-S^2}}\xi, \quad \quad \xi^R=-
\frac {1} {\sqrt{1-S^2}}
C\xi\label{xiLR}
\ee
In this way we have two complementary bases $\xi_n^L$ and $\xi_n^R$.
They are complementary in the sense that
\be
\sum_{n=1}^\infty \Big(\xi_n^L(k)\xi_n^L(k') +\xi_n^R(k)\xi_n^R(k') \Big)=
\delta(k,k')\label{complete2}
\ee
Notice that
\be
\xi_n^R(k)= \xi_n^L(-k), \quad\quad {\rm while}\quad\quad
 \xi_n^R(-k)= \xi_n^L(k)=0.\label{changesign}
\ee
We can project $\xi_n^L$ and $\xi_n^R$ on the
ordinary ${\rm v}_n(k)$ basis of eigenvector of the continuous spectrum,
\cite{RSZ4} (see Appendix) and define the coefficients
\be
b_{nl} =\langle \xi_n^L|{\rm v}_l\rangle,\quad\quad\quad \tilde b_{nl}=
\langle \xi_n^R|{\rm v}_l\rangle\0
\ee

Using the latter we can introduce
\be
\beta_m^\mu = \sum_{l=1}^\infty b_{ml} s_l^\mu,
\quad \quad \tilde\beta_m^\mu =- \sum_{l=1}^\infty \tilde b_{ml}
s_l^\mu\label{beta}
\ee
with the respective hermitian conjugates. The reason for the minus sign in
the second definition above will become clear shortly. These operators
satisfy the algebra
\be
&&[\beta^\mu_m, \beta_n^{\nu\dagger}] = \delta_{m,n} \eta^{\mu\nu}\0\\
&&[\tilde\beta^\mu_m, \tilde\beta_n^{\nu\dagger}]=  \delta_{m,n}
\eta^{\mu\nu}\0
\ee
while all the other commutators vanish.

It must be remarked that the
definition of $\beta_n,\tilde \beta_n$ depends on the $\xi_n$
basis we use. This entails a $O(\infty)$ `gauge' freedom in the choice of
these operators.

These $\beta, \tilde \beta$ operators are natural candidates as closed string
creation and annihilation operators. For the same reason it is natural to
interpret the sliver $|\Xi\rangle$ as the closed string vacuum $|0_c\rangle$.

\section{Properties of the $\beta$ and $\tilde\beta$ operators}

The operators $\beta_n$ and
$\tilde\beta_n$ and their conjugates are
characterized by a Heisenberg algebra isomorphic to the algebra
of closed string creation and annihilation operators. The zero mode
oscillators have not been introduced yet. This will be done in the
following section. Ignoring for the time being the zero modes, in this
section we would like delve into the properties of the
$\beta,\tilde \beta$ operators.

Let us consider the identity
\be
&& \sum_n \beta_n^{\mu\dagger} \tilde \beta_n^{\nu\dagger}\eta_{\mu\nu}= -
\sum_n \langle s^{\mu \dagger}|\xi_n^L\rangle \langle
\xi_n^R|s^{\nu\dagger}\rangle\eta_{\mu\nu} \0\\
&& = \sum_n \langle s^{\mu \dagger}|\xi^L_n\rangle \langle
\xi^L_n| C s^{\nu\dagger}\rangle\eta_{\mu\nu}=
\frac 12 \sum_{k=1}^\infty
s_k^{\mu\dagger} C_{kl} s_l^{\nu\dagger}\eta_{\mu\nu}\0
\ee
The factor of $\frac 12$ comes from the fact that $\xi_n$ is a complete basis
for the left $\xi$'s. We have to consider also the other half made of
$C\xi_n$, which gives the same contribution, see (\ref{complete2}). Hence
the factor of $\frac 12$. The - signs come from the definition (\ref{beta})
and from the property (\ref{changesign}).

Motivated by the above isomorphism we denote $|\Xi\rangle$ by
$|0_c\rangle$ (the `closed string vacuum') when $\beta$ operators are
applied to it. We have the following identity
\be
e^{-\sum_n \beta_n^{\mu\dagger} \tilde
\beta_n^{\nu\dagger}\eta_{\mu\nu}}|0_c\rangle = e^{-\frac 12\sum_{k=1}^\infty
s_k^{\mu\dagger} C_{kl} s_l^{\nu\dagger}\eta_{\mu\nu}}|\Xi\rangle\sim
e^{- \frac 12 \sum_{k=1}^\infty
a_k^{\mu\dagger} C_{kl} a_l^{\nu\dagger}\eta_{\mu\nu}}|0\rangle
\label{boundary}
\ee
where $|0\rangle$ is the original open string vacuum. The last step of the
proof can be found for instance in \cite{aref}, the equality holds up to
a constant.

The LHS is proportional to the boundary state in closed string theory,
the right hand side is the identity state in open string field theory.
The boundary state represents a D25--brane in the closed string language.
The identity state represents absence of interaction in the open string
theory language: the identity state leaves any string state invariant
under star multiplication, and the star multiplication represents
string interaction. An interpretation of this identification
will be presented at the end of section 5.

\subsection{Closed string oscillators as $*$-algebra multiplication operators}

A useful way of defining the $\beta$ and $\tilde\beta$ operators
is as multiplication operators in the $*$-algebra of string
states. This is what we wish to discuss now. At the same time we
would like to introduce the issue of Lorentz covariance. Let us
begin by considering a particular set of half string states,
containing only one closed string excitation, say $l$. A basis for
such states is given by \be
{\Lambda_l}^{\mu_1...\mu_m,\,\nu_1...\nu_n}=\frac
{(-1)^n}{\sqrt{n!m!}}\,
\beta_l^{\mu_1\,\dagger}...\beta_l^{\mu_m\,\dagger}\,
\tilde\beta_l^{\nu_1\,\dagger}...\tilde\beta_l^{\nu_n\,\dagger}\ket\Xi
\label{Lambdal} \ee It is easy to prove that these states form the
following subalgebra (indexes are lowered with the Minkowski
metric) \be {\Lambda_l}^{\mu_1...\mu_n,\,\nu_1...\nu_m}*
\Lambda_{l\rho_1...\rho_p}^{\quad\quad\sigma_1...\sigma_q}=\,
\delta_{mp}\;\hat\delta^{\nu_1...\nu_m}_{\rho_1...\rho_m}\;
{\Lambda_l}^{\mu_1...\mu_n,\,\sigma_1...\sigma_q}\label{L*L} \ee
where we have used the symmetrized delta \be
\hat\delta^{\mu_1...\mu_n}_{\nu_1...\nu_n}=\frac1{n!}\sum_{\sigma(1...n)}
\delta_{\nu_1}^{\mu_{\sigma(1)}}...\;\delta_{\nu_n}^{\mu_{\sigma(n)}}\0
\ee Note that in this new representation the labels (n,m) are
naturally interpreted as two independent (left/right) spin
quantities (number of symmetric indexes). It is easy to prove the
following identities \be &&\beta_l^\mu {\Lambda_l}^{\mu_1...\mu_m,
\nu_1...\nu_n} = \sqrt{m} \,\eta^{\mu(\mu_1}\,
{\Lambda_l}^{\mu_2...\mu_m),\nu_1...
\nu_n}\label{bl1}\\
&&\beta_l^{\mu\dagger} {\Lambda_l}^{\mu_1...\mu_m, \nu_1...\nu_n} =
\sqrt{m+1} \,
{\Lambda_l}^{\mu\mu_1...\mu_m,\nu_1...\nu_n}\label{bl2}\\
&&\tilde\beta_l^\nu {\Lambda_l}^{\mu_1...\mu_m, \nu_1...\nu_n} = \sqrt{n} \,
{\Lambda_l}^{\mu_1...\mu_m,(\nu_1...\nu_{n-1}}\eta^{\nu_n)\nu}\label{bl3}\\
&&\tilde \beta_l^{\nu\dagger} {\Lambda_l}^{\mu_1...\mu_m, \nu_1...\nu_n} =
\sqrt{n+1} \, {\Lambda_l}^{\mu_1...\mu_m,\nu_1...\nu_n\nu}\label{bl4}
\ee
When symmetrizing indexes the normalization is always understood to be
defined, for any tensor $T^{\mu_1...\mu_n}$, by
\be
T^{(\mu_1...\mu_n)} =\frac 1{n!} \sum_\sigma T^{\mu_{\sigma(1)}...
\mu_{\sigma(n)}}\0
\ee

An important subset of these states is formed by the fully traced ones
\be
\Lambda_{l,n} = {{\Lambda_l}^{\mu_1...\mu_n}}_{\mu_1...\mu_n}\0
\ee
It is interesting to consider the sum of all these states
\be\label{completel}
\sum_{ n=0}^\infty\,\Lambda_{l,n}&=&\sum_{n=0}^\infty
\frac{(-1)^n}{n!}\beta_l^{\mu_1\,\dagger}...\,\beta_l^{\mu_{n}\,\dagger}
\tilde\beta_{l\;\mu_1}^{\dagger}...\,
\tilde\beta_{l\;\mu_{n}}^{\dagger}\ket\Xi\0\\
&=&\sum_{n}\frac{(-1)^n}{n!}
\left(\beta_l^\dagger\cdot\tilde\beta_l^\dagger\right)^{n}\ket\Xi
=\,e^{-\beta_l^\dagger\cdot\tilde\beta_l^\dagger}\ket\Xi= \ket
{I_l} \ee The reason for the latter notation is that $\ket {I_l}$
acts as the identity while $*$-multiplying the states
(\ref{Lambdal}).

As in \cite{Furu}, let us introduce the string fields
\be\label{Amul}
&&A^\mu_l =\beta_l^\mu\ket {I_l}= -\tilde\beta_l^{\mu\,\dagger}\ket {I_l}=
\sum_{n=0}^\infty \, \sqrt{n}\, \eta^{\mu\mu_1}
{{\Lambda_l} ^{\mu_2...\mu_n}}_{\mu_1...\mu_n}\0\\
&&A^{\mu,\dagger}_l=\beta_l^{\mu\, \dagger}\ket {I_l}=-
\tilde\beta_l^{\mu}\ket {I_l}= \sum_{n=0}^\infty \,\sqrt{n+1}\,
{{\Lambda_l}^{\mu\mu_1...\mu_n}}_{\mu_1...\mu_n} \ee They
obey\footnote{Note that the first part of (\ref{Amul}) is nothing
but the overlap condition for the Boundary state along the Neumann
directions.}
\be
[A^\mu_l,A^{\nu,\dagger}_r]_*=\delta_{lr}\eta^{\mu\nu}\ket
{I_l}\label{Acomm}
\ee

The action of these string field under left/right
$*$-multiplication on string fields of the type $\ref{Lambdal}$ is
as follows \be
A_l^\mu*{\Lambda_l}^{\mu_1...\mu_m,\,\nu_1...\nu_n}&=
&\sqrt{m}\;\eta^{\mu(\mu_1} \,{\Lambda_l}^{\mu_2...\mu_m),
\,\nu_1...\nu_n}\label{AL}\\
A_{l}^{\mu\,\dagger}*{\Lambda_l}^{\mu_1...\mu_m,\,\nu_1...\nu_n}&=
&\sqrt{m+1}\; {\Lambda_l}^{\mu\mu_1...\mu_m,\,
\nu_1...\nu_n}\label{ADL}\\
{\Lambda_l}^{\mu_1...\mu_m,\,\nu_1...\nu_n}*A_l^{\nu\,\dagger}&=&\sqrt{n}\;
{\Lambda_l}^{\mu_1...\mu_m,\,
(\nu_1...\nu_{n-1}}\,\eta^{\nu_n)\,\nu}\label{AR}\\
{\Lambda_l}^{\mu_1...\mu_m,\,\nu_1...\nu_n}*A_{l}^{\nu}&=&\sqrt{n+1}\;
{\Lambda_l}^{\mu_1...\mu_m,\,\nu_1...\nu_n\nu} \label{ADR}\ee
Comparing (\ref{AL}) with (\ref{bl1}) we see that the left
$*$-multiplication by $A_l^\mu$ corresponds to applying
$\beta_l^\mu$ to the string field. Similarly the left
$*$-multiplication by $A_l^{\mu\, \dagger}$ corresponds to
applying $\beta_l^{\mu\,\dagger}$, the right $*$-multiplication by
$A_l^{\nu\, \dagger}$ corresponds to applying $\tilde\beta_l^\nu$
and the right $*$-multiplication by $A_l^\mu$ corresponds to
applying $\tilde\beta_l^{\nu\,\dagger}$. We recall that $A$ and
$A^\dagger$ should not be confused with creation and annihilation
operators, they are just string fields.

We can now define the half string number operator as follow
\be
N_l=A_{l}^{\mu\dagger}*A_{l,\,\mu}\label{Nl}
\ee
Its left/right action computes the left/right spin of a string field
for given $l$
\be
N_l*{\Lambda_l}^{\mu_1...\mu_n,\,\nu_1...\nu_m}&=&
n\;{\Lambda_l}^{\mu_1...\mu_n,\,\nu_1...\nu_m}\0\\
{{\Lambda_l}}^{\mu_1...\mu_n,\,\nu_1...\nu_m}*N_l&=&
m\;{{\Lambda_l}}^{\mu_1...\mu_n,\,\nu_1...\nu_m}\0
\ee

As we have already noticed, the split left/right structure we are
dealing with is very reminiscent
of the holomorphic/antiholomorphic structure one encounters
in the first quantization of the closed string. This correspondence
can be made more precise. Let us define
\be
{\mathcal A}_{l,\,\mu}=\sqrt{l}\,A_{l,\,\mu}, \quad\quad
{\mathcal A}_{l,\,\mu}^\dagger=\sqrt{l}\,A_{l,\,\mu}^{\dagger}
\label{AlAldagger}
\ee
We get
\be
\left[{\mathcal A}_{l,\,\mu},\,{\mathcal A}_{l,\,\nu}^{\dagger}\right]_*
&=& l\,\eta_{\mu\nu}\,\ket{I_l}\0
\ee

The extension of the above to multiple half string excitations is
straightforward.
We define sequences of natural numbers ${\bf n}=n_1,n_2,...$, where the label
$l$ in $n_l$ corresponds to the oscillator type.
For every type $l$ half string oscillator we will have a collection of
symmetric Lorentz indexes $\mu_1^l,\mu_2^l,...,\mu_{n_l}^l$. Then for any
two sequences ${\bf n}$ and ${\bf m}$ we define (generalizing (\ref{Lambdal}))
the states:
\be\label{Lambdagen}
\Lambda^{\{\mu_1...\mu_{\bf n}\},\,\{\nu_1...\nu_{\bf m}\}}=
\prod_{l,r=1}^{\infty}\frac{(-1)^{m_r}}{\sqrt{n_l!m_r!}}\, \beta_l^{\mu_1^l\,
\dagger}...\beta_l^{\mu_{n_l}^l\,\dagger}\,
\tilde\beta_r^{\nu_1^r\,\dagger}...\tilde\beta_r^{\nu_{m_r}^r\,
\dagger}\ket\Xi
\ee

The complete star algebra is then
\be
\Lambda^{\{\mu_1^l...\mu_{n_l}^l\},\,\{\nu_1^l...\nu_{m_l}^l\}}*
\Lambda_{\{\rho_1^l...\rho_{p_l}^l\}}^{\quad\quad\quad\{\sigma_1^l...
\sigma_{q_l}^l\}}
=\prod_l\,\delta_{m_l,p_l}\;\hat\delta_{\rho_1^l...
\rho_{p_l}^l}^{\nu_1^l...\nu_{m_l}^l}\;\Lambda^{\{\mu_1^l...\mu_{n_l}^l\},
\,\{\sigma_1^l...\sigma_{q_l}^l\}}\label{staralgebra}
\ee
This algebra contains a lot of orthogonal projectors, the simplest ones
being given by Lorentz traces of left/right symmetric states
\be
\tr(\Lambda_{\bf n})&=&
\Lambda_{\{\mu_1^l...\mu_{n_l}^l\}}^{\quad\quad\quad\{\mu_1^l...
\mu_{n_l}^l\}}\0\\
\tr(\Lambda_{\bf n})*\tr(\Lambda_{\bf m})&=&\delta_{\bf nm}\;
\tr(\Lambda_{\bf n})\0
\ee

The sum of these states is the identity string field
\be\label{complete}
\sum_{\bf n=0}^\infty\,\tr(\Lambda_{\bf n})&=&\sum_{n_1}...\sum_{n_\infty}\,
\prod_l\,
\frac{(-1)^{n_l}}{n_l!}\beta_l^{\mu_1^l\,\dagger}...\,\beta_l^{\mu_{n_l}^l\,
\dagger}\tilde\beta_{l\;\mu_1^l}^{\dagger}...\,
\tilde\beta_{l\;\mu_{n_l}^l}^{\dagger}\ket\Xi\0\\
&=&\prod_l\,\sum_{n_l}\frac{(-1)^{n_l}}{n_l!}\left(\beta_l^\dagger\cdot
\tilde\beta_l^\dagger\right)^{n_l}\ket\Xi
=e^{-\sum_l\beta_l^\dagger\cdot\tilde\beta_l^\dagger}\ket\Xi\0\\
&=&e^{-\sum_l\beta_l^\dagger\cdot\, C\beta_l^\dagger}\ket\Xi
=e^{-\frac12 s^\dagger\cdot\, C s^\dagger}\ket\Xi=\ket I
\ee
This sum is nothing but the tensor product of the states $|I_l\rangle$ for
all $l$.
Using the latter identity it is immediate to generalize the definitions of
the states $A$ and $A^\dagger$ by replacing $|I_l\rangle$
with $|I\rangle$. The only change is
\be
\left[{ A}_{l,\,\mu},\,{ A}_{r,\,\nu}^{\dagger}\right]_* &=& \delta_{lr}
\,\eta_{\mu\nu}\,\ket{I}\0
\ee
The left/right action of this operators is
\be
A_l^\mu*\Lambda_r^{\mu_1...\mu_m,\,\nu_1...\nu_n}&=&\delta_{lr}\sqrt{m}\;
\eta^{\mu(\mu_1}\,{\Lambda_l}^{\mu_2...\mu_m),\,\nu_1...\nu_n}\label{Al}\\
A_{l}^{\mu\,\dagger}*\Lambda_r^{\mu_1...\mu_m,\,\nu_1...\nu_n}&=&\delta_{lr}
\sqrt{m+1}
\;{\Lambda_l}^{\mu\mu_1...\mu_m,\,
\nu_1...\nu_n}\label{Adl}\\
\Lambda_r^{\mu_1...\mu_m,\,\nu_1...\nu_n}*A_l^{\nu\,\dagger}&=&
\delta_{lr}\sqrt{n}\;{\Lambda_l}^{\mu_1...\mu_m,\,
(\nu_1...\nu_{n-1}}\,\eta^{\nu_n)\,\nu}\label{Ar}\\
\Lambda_r^{\mu_1...\mu_m,\,\nu_1...\nu_n}*A_{l}^{\nu}&=&\delta_{lr}\sqrt{n+1}
\;{\Lambda_l}^{\mu_1...\mu_m,\,\nu_1...\nu_n\nu} \label{Adr}\ee To
fit the closed string formalism we define the operators ${\mathcal
A}_{l,\,\mu}$ as above, so that \be \left[{\mathcal
A}_{l,\,\mu},\,{\mathcal A}_{r,\,\nu}^{\dagger}\right]_* &=&l\,
\delta_{lr}\,\eta_{\mu\nu}\,\ket{I}\0 \ee Moreover we define the
level ($*$-multiplication) operator \be {\cal N}=
\sum_{l=1}^\infty {\mathcal A}_{l\,\mu}^\dagger * {\mathcal
A}_l^\mu\label{levelop} \ee which counts the level (in the sense
of closed string theory) in the holomorphic sector when acting on
the left and in the antiholomorphic sector when acting on the
right. We can therefore define closed string $\alpha$ operators in
the traditional sense by means of the left/right actions on
generic string states $\psi$ of the type (\ref{Lambdagen}): \be
{\bf \alpha}_{l,\,\mu}\,\psi&=&{\mathcal A}_{l,\,\mu}*\psi\0\\
{\bf \alpha}_{l,\,\mu}^\dagger\,\psi&=&{\mathcal A}^\dagger_{l,\,\mu}*\psi\0\\
{\bf \tilde\alpha}_{l,\,\mu}\,\psi&=&\psi*{\mathcal A}^\dagger_{l,\,\mu}\0\\
{\bf \tilde\alpha}_{l,\,\mu}^\dagger\,\psi&=&\psi*{\mathcal A}_{l,\,\mu}\0
\ee
As expected we get
\be
\left[{\bf \alpha}_{l,\,\mu},{\bf \alpha}_{r,\,\nu}^\dagger\right]&=&
l\,\delta_{lr}\eta_{\mu\nu}\0\\
\left[{\bf\tilde \alpha}_{l,\,\mu},{\bf\tilde \alpha}^\dagger_{r,\,\nu}\right]
&=&l\,\delta_{lr}\eta_{\mu\nu}\0\\
\left[{\bf \alpha},{\bf\tilde \alpha}\right]&=&0\0
\ee

From this one can directly define the holomorphic and anti-holomorphic
Virasoro algebras of the closed string as {\it open} star subalgebras.

To conclude  this discussion, it is worth noting that, even if the above
construction is actually dependent on the ''gauge'' choice of the
oscillators basis $(\beta_l,\tilde\beta_l)$, still one can define
the (closed string) level from a given string field $\psi$ in the $O(\infty)$
star rotation invariant way (see \cite{RSZ3})
\be
&&n_L(\psi) = \frac{\bra{\Omega\psi}\cal N * \psi
\rangle}{\bra{\Omega\psi}\psi\rangle}\label{nL}\\
&& n_R(\psi) = \frac{\bra{\Omega\psi} \psi * \cal N
\rangle}{\bra{\Omega\psi}\psi\rangle} \label{nR}
\ee
where $\Omega$ is the twist transformation which, combined with {\it bpz},
gives

\be
bpz\ket{\Omega\Lambda^{\{\mu_1...\mu_{\bf
n}\},\,\{\nu_1...\nu_{\bf m}\}}} =
\bra{\Lambda^{\{\nu_1...\nu_{\bf m}\},\,\,\{\mu_1...\mu_{\bf
n}\}}} \label{bpzOL}
\ee

See also the comments at the end of section 6 for more on this issue.

\subsection{Representation in terms of Laguerre polynomials}

It is important to clarify the open string nature of the
$\beta,\tilde\beta$ operators.
When applied to the vacuum they turn out to be very well known objects,
which have already made their appearance in SFT. The corresponding states
give rise to an algebra defined by means of Laguerre polynomials.

Let us consider a particular state of the type (\ref{Lambdagen}):
\be
\frac 1 {\sqrt{n!m!}}
(\beta_k^\dagger)^n (-\tilde \beta_l^\dagger)^m |0_c\rangle
\0
\ee
where, for simplicity, we have dropped the Lorentz index $\mu$. Written
out explicitly in open string language the state takes the form
\be
&&(\beta_k^\dagger)^n (-\tilde \beta_l^\dagger)^m |0_c\rangle=
\langle \xi_l C \omega^2 (a^\dagger +S a)\rangle^m
\langle \xi_k \omega^2 (a^\dagger +S a)\rangle^n e^{-\frac 12 a^\dagger S
a^\dagger}|0\rangle \0\\
&& = e^{-\frac 12 a^\dagger S a^\dagger} \langle \xi_l C \omega^2
((1-S^2)a^\dagger +S a)\rangle^m
\langle \xi_k \omega^2 ((1-S^2)a^\dagger +S a)\rangle^n|0\rangle \0\\
&& = e^{-\frac 12 a^\dagger S a^\dagger} \sum_{p=0}^m \left(
\matrix{ m\cr p\cr} \right) \langle \xi_l C a^\dagger\rangle^{m-p}
\langle \xi_l C \omega^2 S a\rangle^p
\langle \xi_k a^\dagger\rangle^n |0\rangle= *\0
\ee
This is due to the fact that the contractions implicit in the reordering of
the above terms all vanish
\be
\langle \xi_k C\frac T{1-T^2} \xi_k \rangle =0,\quad \quad
 \langle \xi_l C \frac T{1-T^2} \xi_l \rangle =0 \0
 \ee
In the next passages we will also need
\be
\langle \xi_k C\frac T{1-T^2} \xi_l \rangle =0\0
\ee
and the definition
\be
\kappa_{kl} = \langle \xi_k \frac T{1-T^2} \xi_l \rangle \label{kappa}
\ee
Proceeding with the algebraic manipulations, and assuming from now on
$n\leq m$,
\be
&&*= e^{-\frac 12 a^\dagger S a^\dagger} \sum_{p=0}^n \left(
\matrix{ m\cr p\cr} \right) \langle \xi_l C a^\dagger\rangle^{m-p}
\langle \xi_k a^\dagger\rangle^{n-p} \frac {n!}{(n-p)!}
(\kappa_{kl})^p|0\rangle\0\\
&&=  e^{-\frac 12 a^\dagger S a^\dagger} \sum_{p=0}^n
\frac{m!n!}{(m-k)!(n-k)!k!}\langle \xi_l C a^\dagger\rangle^{m-n}
 \langle \xi_l C a^\dagger\rangle^{n-p}
\langle \xi_k a^\dagger\rangle^{n-p} \frac {n!}{(n-p)!}
(\kappa_{kl})^p|0\rangle\0\\
&& = (\kappa_{kl})^n\, n!\, Y_l^{m-n} L_n^{m-n} (-\frac
{X_kY_l}{\kappa_{kl}})|\Xi\rangle
\ee
where $X_k = \langle \xi_k a^\dagger\rangle $, $Y_l =
\langle \xi_l C
 a^\dagger\rangle $ and
\be
L_n^{m-n} (z)= \sum_{p=0}^m \left(\matrix{m \cr n-p} \right) \frac
{(-z)^p}{p!},
\quad\quad n\leq m\0
\ee
Therefore, eventually
\be
\frac 1 {\sqrt{n!m!}}
(\beta_k^\dagger)^n (-\tilde \beta_l^\dagger)^m |0_c\rangle=
\sqrt{\frac {n!} {m!}} (\kappa_{kl})^n  Y_l^{m-n} L_n^{m-n} (-\frac
{X_kY_l}{\kappa_{kl}})|\Xi\rangle\label{closedstate}
\ee
For $n=m$ these states have already appeared in the
literature. They have been
interpreted as $D$--brane solutions of vacuum SFT, \cite{BMS3, carlo}.

\section{Closed string states: zero momentum}

Let us return to the main problem. We have seen so far
a correspondence between star algebra operators and closed string creation
and annihilation operators. The relevant question is now: what are the
(open string) string fields that correspond to closed string Fock states
created under the above correspondence? By closed string states we mean both
off-shell and on-shell states. For instance a graviton state
with momentum $k$ in closed string theory is given by
\be
\theta_{\mu\nu}\a_1^{\mu\dagger} \a_1^{\nu\dagger} |0_c,k\rangle
\label{graviton}
\ee
where $|0_c,k\rangle$ is the closed string vacuum with momentum $k$, and
$\theta_{\mu\nu}$ is the polarization.
This state is on-shell when $k^2=0$ and
$\theta_{\mu\nu}k^\nu= \theta_{\mu\nu}k^\mu=0$. When the latter conditions
are not satisfied the graviton is off-shell. Off-shell states
are not so generic as one might think, they must satisfy precise conditions:
they must have definite momentum
(i.e. the holomorphic and antiholomorphic momenta must be equal) and they
must be level--matched. Usually in dealing with closed strings, these two
conditions are so obvious that they are understood, but, as we shall see,
under the correspondence with open strings, they become significant
and select a very precise class of string fields, the projectors. In this
and the next sections we will concentrate on {\it off--shell} closed string
states.
In the present section, to start with, we consider only zero momentum states.
Non-zero momentum states will be introduced in the next section.

It is evident from the above that
there is a correspondence between (zero momentum) states in the Fock space
of the closed string theory and open string fields of the type
(\ref{Lambdagen}). The question is: what are the string fields that correspond
to off--shell states in the closed string theory?

To start with we (formally) define Virasoro generators $L_n, \tilde L_n$
using the $\beta, \tilde\beta$ operators in the usual way. Then using $L_0$
and $\tilde L_0$ we define the mass operator and the level matching condition
by means of
\be
N_L = \sum_{n=1}^\infty n\,\beta_n^\dagger\cdot \beta_n,
\quad\quad  N_R = \sum_{n=1}^\infty n\, \tilde\beta_n^\dagger\cdot
\tilde\beta_n,\label{NRNL}
\ee

Off-shell states are characterized in particular by the condition $N_R=N_L=N$,
where the number $N$ specifies the level of the state. They are in general
combination of monomials of $\beta$ and $ \tilde\beta$ applied to the vacuum
with arbitrary coefficients. The statement we wish to prove is the following:

{\it Closed string Fock space states of given level, satisfying
the level matching condition, can always be decomposed into
combinations of states of the type (\ref{Lambdagen}) that are
$*$-algebra projectors.}  Loosely speaking, level--matched states
of the closed string Fock space come from star algebra projectors.

The relevant states of the Fock space must form representations of the
Lorentz group. However this is not a significant issue here since the
corresponding tensors are saturated with arbitrary polarizations. Let us
write out the states (\ref{Lambdagen}) in the more explicit form
\be
\sim \beta_1^{\mu^1_1\dagger}\beta_1^{\mu^1_2\dagger}\ldots
\beta_1^{\mu^1_{n_1}\dagger}\dots
\beta_l^{\mu^l_1\dagger}\beta_l^{\mu^l_2\dagger}\ldots
\beta_l^{\mu^l_{n_l}\dagger}
\tilde\beta_1^{\nu^1_1\dagger}\tilde\beta_1^{\nu^1_2\dagger}\ldots
\tilde\beta_1^{\nu^1_{m_1}\dagger}\tilde\beta_r^{\nu^r_1\dagger}\ldots
\tilde\beta_r^{\mu^r_2\dagger}\ldots
\tilde\beta_r^{\mu^r_{n_r}\dagger}|\Xi\rangle\label{fam2}
\ee
We can rewrite this in the form
\be
&&\sim (\beta_1^{0\dagger})^{n_{0,1}}(\beta_1^{1\dagger})^{n_{1,1}}\ldots
(\beta_1^{25\dagger})^{n_{25,1}}(\beta_2^{0\dagger})^{n_{0,2}}\ldots
(\beta_2^{25\dagger})^{n_{25,2}}\ldots(\beta_l^{0\dagger})^{n_{0,l}}\ldots
(\beta_l^{25\dagger})^{n_{25,l}}\0\\
&&~~~~~~~~~~~(\tilde\beta_1^{0\dagger})^{m_{0,1}}
\ldots(\tilde\beta_1^{25\dagger})^{m_{25,1}}\ldots
(\tilde\beta_r^{0\dagger})^{m_{0,r}}\ldots
(\tilde \beta_r^{25\dagger})^{m_{25,r}}|\Xi\rangle
\label{fam1}
\ee
It is evident that the first family of terms coincides with the second one
provided that $\sum_\mu n_{\mu,i} =n_i$, $i=1,\ldots, l$ and
$\sum_\nu m_{\nu,j} =m_j$ with $j=1,\ldots,r$.
In the sequel we preferably use the form (\ref{fam1}).

A state belonging to the closed string Hilbert space, even though it is not
on--shell, satisfies the level matching condition, i.e.
$N_L= \sum_{i=1}^l in_i$ and
$N_R=\sum_{j=1}^r j m_j$ coincide, $N_L=N_R=N$. It is a combination with
arbitrary coefficients of all
the states of the type (\ref{Lambdagen}) satisfying this condition.
However, to start with, let us ignore the complication
of the Lorentz indexes and drop the index $\mu$ altogether
(i.e. we pretend there is only one space--time direction).
In this case the states can be fully identified by the symbols
$\Lambda_{\bf n,\bf m}$, because they are completely specified if we know
the two sequences ${\bf n}$ and ${\bf m}$
\be
\Lambda_{\bf n,\bf m}= \prod_{l,r=1}^{\infty}
\frac{(-1)^{m_r}}{\sqrt{n_l!m_r!}}\, (\beta_l^\dagger)^{n_l}\,
(\tilde\beta_r^{\dagger})^{m_r}\ket\Xi \label{Lambdanm}
\ee
One has
\be
\Lambda_{\bf n,\bf m}*\Lambda_{\bf p,\bf q}=
\delta_{\bf m,\bf p} \Lambda_{\bf n,\bf q}\label{Lnm*Lpq}
\ee
where $\delta_{\bf m,\bf p} =\prod_{l,r}\delta(m_l,p_r) $. An off--shell
closed string state
will therefore be represented by a superposition of states
$\Lambda_{\n,\m}$ in the closed string Fock space. Setting $N_L=N_R=N$, there
will be a leading
state with $\n=\m= (0,...,0,1,0,....)$ with one single non-vanishing entry
equal to $1$
in the $N$--th position (which corresponds to one single operator
$\beta_N^\dagger$ and one single operator $\tilde \beta_N^\dagger$ of highest
order applied to the vacuum), which we denote simply by $\Lambda_{N,N}$. We
refer to the full set of
states as the family ${\cal F}_{(N,N)}$. We have now the problem of
dealing with all the other states $\Lambda_{\n,\m}$ in the family different
from $\Lambda_{N,N}$. To this end we recall that any sequence
$\n$ naturally represents a partition of $N$ (i.e. $(n_1, n_2,...)$
is read as the partition such that $\sum_i in_i=N$, the sequence
corresponding to the partition that contaisn only $N$ itself
will be denoted by $\n_N$).
In order to be able to deal with all these possibilities, we introduce a
partial ordering among the sequences $\n$: we say that $\n\geq  \n'$ iff
the rightmost nonzero $n_i$ and $n'_{i'}$ in $\n$ and $\n'$, respectively,
are such that $i\geq i'$ and, if $i=i'$, $n_i \geq n'_{i}$, and, if also
$n_i=n'_i$ the second rightmost numbers in $\n$ and $\n'$ are to be considered,
and so on. Among the states of the family let us
pick those of the form $\Lambda_{\n,\n}$, among which is $\Lambda_{N,N}$,
and call them {\it principal}. All the other states are {\it descendants}.
Given a principal state we can define a subfamily as follows: it contains
the principal state $\Lambda_{\n,\n}$ as well as $\Lambda_{\n,\n_d}$ and
$\Lambda_{\n_d,\n}$, where $\n_d$ is any sequence $\leq \n$
but not identical to it. This will be
referred to as the $(\n,\n)$ subfamily. All the subfamilies are naturally
ordered, the highest one being $(N,N)$.

Let us start from $\Lambda_{N,N}$ and the relative subfamily. On the basis
of (\ref{staralgebra}), $\Lambda_{N,N}$ defines a projector. Then we consider
$\Lambda_{N,\m_d}$
where $\m_d$ is any $\m_d \leq \m_N$ but not identical to $\m_N$ (for instance
one $\m_d$ is $(1,0,...,0,1,0,...)$ where the nonzero entries
are in the first and $N-1$--th position, which corresponds to the product
$\tilde \beta_1^\dagger \tilde \beta_{N-1}^\dagger$).
$\Lambda_{N,\m_d}$  is not a star--projector, but the sum
$\Lambda_{N,N}+a\Lambda_{N,\m_d}$ is, for any arbitrary
constant $a$. Indeed, using (\ref{Lnm*Lpq}), one gets
\be
(\Lambda_{N,N}+a\Lambda_{N,\m_d})*(\Lambda_{N,N}+a\Lambda_{N,\m_d}) =
(\Lambda_{N,N}+a\Lambda_{N,\m_d})\0
\ee

 If we take a combination of the
corresponding closed string states with arbitrary coefficient we see that
the combination will contain, beside $\Lambda_{N,N}$, any state
$\Lambda_{N,\m_d}$ with an arbitrary
coefficient in front of it. The same can be said of the states
$\Lambda_{\n_d, N}$, with $\n_d<n_N$. Now we can pass to the second subfamily
and play the same game, the role of $\Lambda_{N,N}$ being played by the
relevant $\Lambda_{\n,\n}$. We can do the same for any subfamily and therefore
exhaust the full set of states, showing that each one of them can be inserted
into a distinct projector. We can claim therefore that each distinct term
in the family ${\cal F}_{(N,N)}$ corresponds to a distinct projector.

Extending the previous proof to the states (\ref{Lambdagen}) is
straightforward. Remembering (\ref{fam1}) the latter can be written in
terms of a multi--sequences $\BN= ({\bf n}_0,{\bf n}_1,..., {\bf n}_{D-1})$.
\be
\Lambda_{\bf N,\bf M}\,=\,
\prod_{\mu=0}^{D-1} \prod_{l,r} \frac {(-1)^{n_{\mu,r}}}{\sqrt{n_{\mu,l}!
m_{\mu,r}!}} (\beta_l^{\mu\dagger})^{n_{\mu,l}}
(\tilde \beta_r^{\mu\dagger})^{n_{\mu,r}}\ket\Xi
\label{LambdaNM}
\ee
In this new notation (\ref{staralgebra}) becomes
\be
\Lambda_{\bf N,\bf M}*\Lambda_{\bf P,\bf Q}= \delta_{\bf M \bf P}
\Lambda_{\bf N,\bf Q}\label{LMN*LPQ}
\ee
where $\delta_{\bf M \bf P} = \prod_\mu \prod_{l,r} \delta (m_{\mu,l},
p_{\mu,r})$.

What one has to do is first of all to define an ordering for the
multi--sequences $\BN$. This is easy to accomplish, starting for instance
from the ordering of the $\n_0$ sequence, then looking at the ordering
of $\n_1$ and so on, and then proceeding as above. Moreover
$N_L = \sum_\mu \sum _i i n_{\mu,i}$ and $N_R = \sum _\mu \sum_j jm_{\mu,j}$.
One can define families
characterized by the properties of containing all the states with $N_L=N_R=N$.
Likewise one can define principal states (having $\BN=\BM$) and subfamilies
as above, and extend the previous proof to such generalized families.

\section{Closed string states: the momentum eigenfunction}

Every closed string state is constructed by tensoring a Fock space state
with a momentum eigenfunction, which, in the coordinate representation,
is the plane wave $e^{ikx}$. The momentum $k$ comes in equal parts from the
left and the right-handed sectors. The purpose of this section is to explain
where this factor comes from in the open-closed correspondence of the previous
sections. Once more we shall see that the origin of this factor is a star
algebra projector.

To start with we remark that in the previous sections all developments were
based on the sliver projector, which is translationally invariant in all
directions. If we want to find a momentum dependence we have therefore
to start from projectors
that are not translationally invariant.
To this end we will use the {\it lump} projector. Let us recall what it is.
The lump projector was engineered to represent a lower dimensional brane
(Dk-brane) in VSFT, therefore it has ($25-k$)
transverse space directions along which translational
invariance is broken.
Accordingly one splits the three string vertex into the tensor product of
the perpendicular part and the parallel part
\be
|V_3\rangle = |V_{3,\perp}\rangle \, \otimes\,|V_{3,_\|}\rangle\label{split}
\ee
The parallel part is the same as in the sliver case while the
perpendicular part is modified as follows.
Following \cite{RSZ2}, we denote by $x^{\bar \mu},p^{\bar \mu}$,
${\bar \mu}=1,...,k$ the
coordinates and momenta in the transverse directions and introduce the
canonical zero modes oscillators
\be
a_0^{(r){\bar \mu}} = \frac 12 \sqrt b \hat p^{(r){\bar \mu}}
- i\frac {1}{\sqrt b} \hat x^{(r){\bar \mu}},
\quad\quad
a_0^{(r){\bar \mu}\dagger} = \frac 12 \sqrt b \hat p^{(r){\bar \mu}} +
i\frac {1}{\sqrt b}\hat x^{(r){\bar \mu}}, \label{osc}
\ee
where $b$ is a free parameter.
Denoting by $|\Omega_{b}\rangle$ the oscillator vacuum
(\,$a_0^{\bar \mu}|\Omega_{b}\rangle=0$\,),
in this new basis the three string vertex is given by
\be
|V_{3,\perp}\rangle'= K\, e^{-E'}|\Omega_b\rangle\label{V3'}
\ee
$K$ being a suitable constant and
\be
E'= \frac 12 \sum_{r,s=1}^3 \sum_{M,N\geq 0} a_M^{(r){\bar \mu}\dagger}
V_{MN}^{'rs} a_N^{(s){\bar \nu}\dagger}\eta_{{\bar \mu}{\bar \nu}}\label{E'}
\ee
where $M,N$ denote the couple of indexes $\{0,m\}$ and $\{0,n\}$,
respectively.
The coefficients $V_{MN}^{'rs}$ are given in Appendix B of \cite{RSZ2}.
The new Neumann coefficients matrices $V^{'rs}$ satisfy the same relations as
the $V^{rs}$ ones. In particular one can introduce the matrices $X^{'rs}=
C V^{'rs}$, where $C_{NM}=(-1)^N\, \delta_{NM}$.
The lump projector $|\Xi'_k\rangle$ has the form (\ref{Xi}) with $S$ along
the parallel directions, while $|0\rangle$ is replaced by $|\Omega_b\rangle$
and  $S$ is replaced by $S'$ along the perpendicular ones.
Here $S'=CT'$ and $T'$ has the same form as $T$ eq.(\ref{sliver}) with
$X$ replaced by $X'$. The normalization constant $\N'$ is defined in a way
analogous to $\N$. The diagonal representation of $X^{'rs}$ is summarized in
Appendix.

We now repeat the same steps as in section 2 in order to define the operators
$\beta_N$ and $\tilde \beta_N$. We are of course interested in particular
in the zero mode. Let us consider a lump projector $|\Xi'\rangle$ and
concentrate on a
transverse direction, say $\mu$. We introduce, in a way analogous to section
2, left and right Fock space projectors $\rho_L'$ and $\rho_R'$, with the same
properties as $\rho_L$ and $\rho_R$, which will not be repeated here. These
operators can be diagonalized (see Appendix). Differently from the sliver
case here we have both a continuous and discrete spectrum. The continuous
spectrum is spanned by a real number $k$, $-\infty<k<+\infty$. The discrete
spectrum can be written in terms of a positive real number $\eta$ and by
$-\eta$ ($\eta$ is related to the parameter $b$, see Appendix). The
corresponding eigenvectors are denoted $V_N(k), V_N(\eta), V_N(-\eta)$.
Their completeness relation can be found in eq.(\ref{completelump}).
Using this basis, $S'$ and $\omega'=1/\sqrt{1-T^{'2}}$, we write down the
analog of formula (\ref{Bog}). The operators ${s'_M}^\mu$ satisfy the
Heisenberg algebra
\be
[{s'_M}^\mu,{s'_N}^{\nu\dagger}] = \delta_{MN} \eta^{\mu\nu}\label{commMN}
\ee
and annihilate the lump projector $|\Xi'\rangle$.

In the diagonal representation $\rho_L'$ and $\rho_R'$ take the following
form:
\be
\rho_R'= \int_0^\infty |k\rangle dk\bra k + |\eta\rangle \bra \eta,\quad\quad
\rho_L'= \int_{-\infty}^0 |k\rangle dk\bra k + |-\eta\rangle \bra{-\eta}\0
\ee
where $\ket k, \ket \eta$ and $\ket{-\eta}$ form a basis such that
$\bra k V_N\rangle= V_N(k), \bra \eta V_N\rangle=V_N(\eta)$
and $ \bra {-\eta} V_N\rangle=V_N(-\eta)$.

In analogy with what we did in section 2 in the sliver case, we define now
vectors $\xi'$ such that $\rho_L' \xi'=\xi'$ and $\rho_R'\xi'=0$.
There exists a complete basis of $\xi'_N$ ($N=0,1,2,...$) that satisfy these
conditions and are orthonormal in the sense that
\be
\langle \xi'_N|\frac 1{1-{T'}^2} |\xi'_M\rangle = \delta_{NM}\label{normal'}
\ee
Then we define
\be
{\xi'_N}^{L} = \omega'\xi'_N, \quad \quad {\xi'_N}^R= \omega'
C\xi'_N\label{xi'LR}
\ee
When projected on the continuous basis $\ket k$ and the discrete one
$\ket\eta$, $\ket {-\eta}$, they give rise to a vector of functions and
numbers ${\xi'_N}^{L}(k)$, ${\xi'_N}^{L}(-\eta)$ and ${\xi'_N}^{R}(k)$,
respectively, ${\xi'_N}^{R}(\eta)$, which satisfy the orthogonality relations
\be
&&\sum_{N=0}^\infty \Big({\xi'_N}^L(k){\xi'_N}^L(k') +
{\xi'_N}^R(k){\xi'_N}^R(k') \Big)=
\delta(k,k')\label{complete2'}\\
&&\sum_{N=0}^\infty \Big({\xi'_N}^L(\eta){\xi'_N}^L(\eta)+
{\xi'_N}^R(\eta){\xi'_N}^R(\eta)
\Big)= 1\label{completedis'}
\ee
For later purposes it is convenient to choose the basis in such a way that
\be
{\xi'_0}^L(-k) ={\xi'_0}^R(k)=0,\quad\quad {\xi'_n}^R(\eta)=
{\xi'_n}^L(-\eta)=0,\quad k>0,\quad n=1,2,...\label{choice}
\ee
This will allow us to separate the continuous from the discrete
spectrum--dependent objects.

Now, in analogy with section 2, we define the coefficients
\be
b'_{NM} = \bra {{\xi'_N}^L} V_M\rangle, \quad\quad \tilde b'_{NM} = \bra
{{\xi'_N}^R} V_M\rangle
\label{b'}
\ee
and the operators
\be
\beta_N^\mu= \sum_{M=0}^\infty b'_{NM} {s'_M}^{\mu},\quad\quad
\tilde\beta_N^\mu= -\sum_{M=0}^\infty \tilde b'_{NM} {s'_M}^{\mu}\label{beta'}
\ee
Needless to say they satisfy the algebra
\be
[\beta_M^\mu,\beta_N^{\nu\dagger}]= \eta^{\mu\nu} \delta_{MN},\quad\quad
 [\tilde\beta_M^\mu,\tilde\beta_N^{\nu\dagger}]= \eta^{\mu\nu}
 \delta_{MN},\label{cancomm}
\ee
while the other commutators vanish. Here $\mu,\nu$ are any two transverse
directions. We remark that we have dropped the prime from the $\beta$'s,
in order to use a uniform notation for the closed string operators. However
it should be kept in mind that the $\beta_n,\tilde \beta_n$ operators are
different from those defined in section 2. We will return to this point
later on.

We are now ready to discuss the momentum eigenstates. To start with let us
define the state
\be
|p,q\rangle = \frac 1K\sqrt{\frac b{2\pi}} e^{-\frac b{4} (p^2+q^2)+
{\sqrt{b}} (q\beta_0^\dagger+p \tilde \beta_0^\dagger)-
\frac 12(\beta_0^{\dagger\, 2} +
\tilde \beta_0^{\dagger\, 2})}|0'_c\rangle\label{pq}
\ee
where $p$ and $q$ are real numbers, $K$ is the constant that appear in
eq.(\ref{V3'}), and $|0'_c\rangle$ stands for the lump $|\Xi'\rangle$.
For notational simplicity we drop Lorentz indexes. They can be
straightforwardly reinserted when needed. We remark the $\beta_0$ and
$\tilde \beta_0$ are not self-adjoint, therefore they cannot be interpreted
as momenta, not even as half--momenta. We define the selfadjoint
half--momenta operators as
\be
\hat q =\frac 1{2\sqrt b}( \beta_0+\beta_0^\dagger),\quad\quad \hat p =
\frac 1{2\sqrt b} (\tilde \beta_0+\tilde\beta_0^\dagger)
\label{hatpq}
\ee
It is easy to verify that the states (\ref{pq}) satisfy
\be
\hat p |p,q\rangle= \frac p2|p,q\rangle, \quad\quad \hat q |p,q\rangle=
\frac q2|p,q\rangle\0
\ee
Now we are going to compute the star product of two such $|p,q\rangle$ states.
The formula for the star product is considerably simplified if
(like in our case) the vacuum state is $|\Xi'\rangle$ instead of the ordinary
open string vacuum. In fact the vertex can be written in the following
shorthand form
\be
\bra {V_3'} =K\;{}_{123}\bra{\Xi'}e^{-\frac 12 s^{(a)}C \hat V^{ab}C
s^{(b)}}\label{3strings}
\ee
where $a,b=1,2,3$ label the three strings and
\be
C\hat V= \left( \matrix {0&\rho_L'&\rho_R'\cr
                  \rho_R'&0&\rho_L'\cr
           \rho_L'&\rho_R'&0\cr}  \right)\label{CV}
\ee
Another prescription one must introduce is the $bpz$ transformation for
the zero modes\footnote{The + sign in the RHS of (\ref{bpz}) is to be
traced back to the - sign in the second eq.(\ref{beta'}).}
\be
bpz(\beta_0) = +\tilde \beta_0^\dagger\label{bpz}
\ee
The star product of two states like (\ref{pq}) can now be straightforwardly
computed, because, due to the choice of basis (\ref{choice}), the zero mode
calculation decouples from the rest. The only caution one must exercise is
introducing a regulator since a naive calculation would bring about infinite
factors. This is easily accomplished by multiplying the term
$(\beta_0^{\dagger\, 2} +\tilde \beta_0^{\dagger\, 2})$ in the exponent of
(\ref{pq}) by a parameter $\epsilon$ and eventually taking the limit
$\epsilon\to 1$.
The result is as follows
\be
|p_1,q_1\rangle * |p_2,q_2\rangle = \lim_{\epsilon \to 1}\;C(\epsilon,
q_1,p_2)\,|p_1,q_2\rangle\0
\ee
where
\be
C(\epsilon,q_1,p_2)= \frac 12 \sqrt{\frac{b}{\pi(1-\epsilon)}}\,
e^{-\frac{b(q_1-p_2)^2}{4(1-\epsilon)}}\0
\ee
The limit for $\epsilon\to 1$ of this expression is $\delta(q_1-p_2)$.
Therefore
\be
|p_1,q_1\rangle * |p_2,q_2\rangle =  \delta({q_1-p_2})
|p_1,q_2\rangle\label{pq1pq2}
\ee
This equation is clearly the natural generalization of equations like
(\ref{L*L}) and
(\ref{Lnm*Lpq}), when continuous parameters are involved (instead of discrete
indexes). For this reason we say that $|p,p\rangle$ is a star algebra
projector (by slightly extending this notion). We remark that this happens
when the left half--momentum is equal to the right half--momentum.

We can therefore improve our description of the closed string states, by
giving them a nonzero momentum in the transverse directions: we tensor the
states discussed in the previous sections
(constructed as in the previous sections, but out of $\beta_n^{\mu \dagger}$
and $\tilde\beta_n^{\mu\dagger}$ given by eq.(\ref{beta'}))
with momentum eigenstates $|p,p\rangle$. The resulting
state will have transverse momentum $p$, which is the eigenvalue of
$\frac 1{2\sqrt b}
(\beta_0^\mu+\beta_0^{\mu\dagger}+\tilde\beta_0^\mu +\tilde
\beta_0^{\mu\dagger})$.

What about the longitudinal momentum? In the
longitudinal directions the star product is determined by the three strings
coefficients $V_{nm}^{ab}$ and setting the momenta to zero (instead of
integrating over them, see \cite{RSZ2}), while the corresponding (zero
momentum off--shell) closed string states have been introduced in section 4.
Generating a momentum eigenfunction in this context is impossible. Some
kind of modification has to be introduced. This would require a rather long
digression. Since longitudinal momenta do not enter in what follows
we postpone dealing with this issue to another occasion.

It is instructive to complete this subject by giving further properties
of the momentum eigenstates and their conjugates. To start with we get the
following $bpz$ product
\be
\langle p,p |q,q\rangle = \frac 1{K^2} \sqrt{\frac b{2\pi}}
\delta(p+q)\label{bpzprod}
\ee
from which we see that the normalization of $|p,p\rangle$ as a star projector
differs from the normalization as a wavefunction.

In order to introduce the coordinate eigenstates let us define
\be
|x,y\rangle = \frac 1K \sqrt{\frac 2{b\pi}} \;e^{-\frac 12 (x^2+y^2) -\frac
{2i}{\sqrt b} (y\beta_0^\dagger-x\tilde\beta_0^\dagger)+\frac 12
(\beta_0^{\dagger\,2}+\tilde \beta_0^{\dagger\, 2})}\,|0_c'\rangle\label{xy}
\ee
The star algebra yields
\be
|x_1,y_1\rangle * |x_2,y_2\rangle = \delta(y_1-x_2)
|x_1,y_2\rangle\label{xy1*xy2}
\ee
Like before, $|x,x\rangle$ can be interpreted as a star algebra projector.
However the position operators are
\be
\hat x =i\frac {\sqrt b}2 (\tilde\beta_0-\tilde\beta_0^\dagger),\quad\quad
\hat y= i\frac {\sqrt b}2  (\beta_0-\beta_0^\dagger), \0
\ee
so that
\be
[\hat x+\hat y, \hat p +\hat q]= i,\0
\ee
as it must be. But we get
\be
\hat x |x,y\rangle = -x|x,y\rangle,\quad\quad \hat y |x,y\rangle =
y|x,y\rangle\0
\ee
So that the position eigenvalue, that is the eigenvalue of $\hat x+\hat y$,
is $z=y-x$. Therefore $|x,x\rangle$ has position $z=0$. In order to get
something meaningful we should choose, as position eigenstate $|-x,x\rangle$.
This is confirmed by the following fact.
When we contract $|x,y\rangle$ with the $|p,q\rangle$ we find
\be
\langle p,q|x,y\rangle \sim e^{i(-px+qy)}\0
\ee
When $p=q$ and $x=y$ this becomes a constant.
Therefore $|x,x\rangle$ cannot be interpreted as a position eigenstate.
On the contrary $|-x,x\rangle$ works very well as a position eigenstate. But
it is not a star projector.

This fact seems to translate at the level of star
algebra the quantum impossibility of simultaneously describing coordinate
and momentum.

\section{The boundary state in the transverse directions}

It is very instructive to redo the computation we did at the beginning of
section 3 for transverse directions. Let $ij$ denote transverse
directions and let us consider the identity
\be
&& \sum_n \beta_n^{i\dagger} \tilde \beta_n^{j\dagger}\eta_{ij}= -
\sum_n \langle s^{'i \dagger}|{\xi'_n}^L\rangle \langle
{\xi'_n}^R|s^{'j\dagger}\rangle\eta_{ij} \0\\
&& = -\sum_n \langle s^{'i \dagger}|{\xi'_n}^L\rangle \langle
{\xi'_n}^L| C s^{'j\dagger}\rangle\eta_{ij}=-
\frac 12 \sum_{k=1}^\infty
{s'_k}^{i\dagger} C_{kl} {s'_l}^{j\dagger}\eta_{ij}\0
\ee
The factor of $\frac 12$ comes from the fact that $\xi'_n$ is a complete basis
for the left $\xi'$'s, as far as the continuous spectrum is concerned
(see (\ref{choice})). We have to consider also the other half made of
$C\xi'_n$, which gives the same contribution, see (\ref{complete2}). Hence
the factor of $\frac 12$. The -- sign come from the definition (\ref{beta'}).
This is not compensated anymore now by the twist properties of the basis
since
\be
\xi^{'R}= C\xi^{'L},\label{changesign'}
\ee
which in turn descends from the sign change of eq.(\ref{difference}) in
Appendix in passing from the `sliver basis' to the `lump basis'.

For the transverse directions we have therefore
the following identity
\be
e^{\sum_n \beta_n^{i\dagger} \tilde
\beta_n^{j\dagger}\eta_{ij}}|0_c\rangle = e^{-\frac 12\sum_{k=1}^\infty
{s'_k}^{i\dagger} C_{kl} {s'_l}^{j\dagger}\eta_{ij}}|\Xi\rangle\sim
e^{ -\frac 12 \sum_{k=1}^\infty
a_k^{i\dagger} C_{kl} a_l^{j\dagger}\eta_{ij}}|0\rangle\label{boundary'}
\ee
where $|0\rangle$ is the original open string vacuum.

Suppose we have Dk--brane in closed string theory, i.e. we have
$25-k$ transverse directions and $k+1$ parallel ones (including
time). Then the oscillator part of the corresponding boundary state in closed
string theory is
the tensor product of a factor like the LHS of eq.(\ref{boundary})
and a factor given by the LHS of the above eq.(\ref{boundary'}). As one can
see the RHS of the two equations takes the same form. This miracle has to be
traced back to the twist properties of the `sliver basis' and the
`lump basis'.

The identification (\ref{boundary'}) generalizes the corresponding result in
section 3. But we are now in a position to offer an interpretation of it.
The LHS is proportional to the boundary state in closed string theory,
the right hand side is the identity state in open string field theory.
The boundary state represents a Dk--brane in the closed string language.
The identity state represents absence of interaction in the open string field
theory language. We can
interpret the above equality in the following way: closed strings are
reflected by the Dk--brane (they feel it). Open string live on the Dk--brane,
therefore they perceive the corresponding state as an identity state
(they don't feel it).

At this stage it is also clear that one cannot speak about closed string
states
in absolute generality but only with respect to a given background. The
closed string states we have introduced are the ones that interact with
the open string excitations of a given D--brane, which is manifest
in the structure of the vacuum they act upon.

The elements brought forth in this section are evidence in favor of
our identification of closed string modes with open string star algebra
projectors. In particular the above mentioned automatic change in boundary
conditions can hardly be a mere accident.

\subsection{Closed string exchange between two boundary states}

As a consistency check of the identification made above,
in this section we would like to reproduce the well known
computation describing the closed string exchange between two
D--brane, by explicitly converting closed string oscillators into
star algebra inner operators. To start with let us recall the basic
result (see for instance \cite{divecchia})\footnote{We disregard
the ghost contribution (which modifies $D\to D-2$ in the last parenthesis).}
\be\label{tensionbound}
\bra{B(0)}\hat D\ket{B(y^i)}
=V_{p+1}\, N_p\, T_p \int_0^\infty dt\, t^{-\frac{D-p-1}{2}}
e^{-\frac{y^2}{2\pi\alpha't}}\,e^{2\pi t}\,\prod_{n=1}^{\infty}
\left(\frac{1}{1-e^{-2\pi nt}}\right)^D
\ee
where $T_p$ is the Dp--brane tension and $N_p$ is a proportionality constant.
Introducing the matrix $B_{\mu\nu}= (\eta_{\alpha\beta},-\delta_{ij})$,
the (matter part of the) boundary state at transverse position $y^i$ is given
by
\be
\ket{B(y^i)}=\frac{T_p}{2}\;\exp\,\left(-\sum_{n=1}^{\infty}\beta^\dagger_n\,
\cdot B\,\cdot\, \tilde{\beta}^\dagger_n\right)\,\delta(\hat x^i- y^i)
\ket{p^\alpha=0}\label{Byi}
\ee
and the closed string propagator is
\be
\hat D=\frac{\alpha'}{4\pi}\int_{|z|\leq1}\,\frac{d^2z}{|z|^2}\,z^{L_0-1}\,
{\bar z}^{{\tilde L}_0-1}\0
\ee

Let us first compute the contribution from nonzero modes by using
the dictionary introduced above. The nonzero mode part is given by
\be\label{propnonzero}
\bra0\exp\,\left(-\sum_{n=1}^{\infty}\beta_n\,\cdot B\,\cdot\,
\tilde{\beta}_n\right)\,z^N\,{\bar z}^{\tilde N}
\exp\,\left(-\sum_{n=1}^{\infty}\beta^\dagger_n\,\cdot B\,\cdot\,
\tilde{\beta}^\dagger_n\right)\ket{0}=\prod_{n=1}^{\infty}\,
\left(\frac{1}{1-|z|^{2 n}}\right)^D
\ee
where $N$ and $\tilde N$ are the usual closed string holomorphic and
anti--holomorphic level operators. In OSFT language these two operators are
given by the left/right action of the operator defined in (\ref{levelop}),
explicitly
\be
N\ket\psi=\ket{{\cal N} * \psi}, \quad\quad
\tilde N\ket{\psi}=\ket{ \psi*{\cal N}}\0
\ee
As we have already shown, the oscillator part of the boundary state gets
mapped to the identity string field in the  non--zero mode sector.
Hence the expression (\ref{propnonzero}) is proportional to
\be
\bra{I}z^{\cal N}*I*\bar z^{\cal N}\rangle=\bra{z^{\cal N}}
\bar z^{\cal N}\rangle\0
\ee
The string field level operator decomposes into the sum of all half string
levels
\be
{\cal N}=\sum_{l=1}^\infty {\cal N}_l\0
\ee

Let's now write $z^{\N}$ in terms of the building blocks $\Lambda_{nm}$.
\be
z^\N=z^{\sum_l \N_l}=\otimes_{l}z^{\hat\N_l}\0
\ee
Notice that the vacuum state (the sliver or the lump, according to the
direction) is actually the tensor product of
the vacua for the oscillators $(\beta_l,\tilde\beta_l)$
\be
\Xi=\otimes_l\Xi_l\0
\ee
Accordingly the star product also factorizes
\be
*=\otimes_l\;(*_l)\0
\ee
This factorization obviously extends to string fields containing just on
$l$--type of operators. To be precise
\be
\N_l&=&\hat\N_l\otimes_{r\neq l}\hat\Xi_r\0\\
\Lambda^{(l)}_{n_l,n_l}&=&\hat\Lambda^{(l)}_{n_l,n_l}\otimes_{r\neq l}
\hat\Xi_r\0
\ee
With the above understanding it is immediate to see that
\be
\hat\N_l=\sum_{n_l=0}^{\infty}\,l\,n_l\,\hat\Lambda^{(l)}_{n_l,n_l}.\0
\ee
Now we explicitly get
\be
z^\N=\otimes_{l}\left(\sum_{n_l=0}^{\infty}\,z^{l\,n_l}\,
\hat\Lambda^{(l)}_{n_l,n_l}\right)\0
\ee
with an analogous result for $\bar z^\N$.
We can finally write (considering just 1 space-time dimension)
\be
\bra{z^{\cal N}}\bar z^{\cal N}\rangle^{(D=1)}&=&\otimes_{l}
\left(\sum_{n_l=0}^{\infty}\,z^{l\,n_l}\,
\bra{\hat\Lambda^{(l)}_{n_l,n_l}} \right)
\otimes_{l'}\left(\sum_{m_l'=0}^{\infty}\,\bar z^{l'\,m_l'}\,
\ket{\hat\Lambda^{(l')}_{m_l',m_l'}}\right)\0\\
&=&\otimes_l\left(\sum_{n_l}\,|z|^{2l\,n_l}\bra{\hat\Xi_l}\hat\Xi_l\rangle
\right)\0\\
&=&\prod_l\left(\frac1{1-|z|^{2l}}\right)\,\otimes_l\,\bra{\hat\Xi_l}
\hat\Xi_l\rangle=\prod_l\left(\frac1{1-|z|^{2l}}\right)\,
\bra{\Xi}\Xi\rangle^{(D=1)}
\ee

Taking into account the  total number of dimensions we finally get
\be\label{finalstar}
\bra{z^{\cal N}}\bar z^{\cal N}\rangle=
\bra{\Xi_p} \Xi_p\rangle\,\prod_{l=1}^{\infty}\,
\left(\frac{1}{1-|z|^{2 l}}\right)^D
\ee
Here we denote by $\Xi_p$ the sliver in the $p+1$ longitudinal direction,
tensored with the lump on the remaining $D-p-1$'s.

Next let us turn to the zero mode part of (\ref{tensionbound}).
To reproduce it
in the open string language let us concentrate on the zero mode part of
(\ref{Byi}): $\delta(\hat x^i- y^i) \ket{p_\perp =0}$  (we disregard the
longitudinal part, which is trivial). Therefore we have to represent,
in the open string language, such states as
\be
 \delta(\hat x- y) \ket{p}= \frac 1{2\pi} \int dq\,e^{iq(\hat x-y)}
\ket{p}\label{xp}
\ee
where, once again, we have dropped all Lorentz indexes and concentrated on
a single transverse direction. Now we represent $\ket {p}$ by means
of the star projector $\ket{p,p}$ introduced in section 5 and $\hat x$
by the operator
$i\frac {\sqrt b}2 (\tilde\beta_0-\tilde\beta_0^\dagger+\beta_0-
\beta_0^\dagger)$, considered in the same section. Then it is easy
to verify that
\be
e^{iq(\hat x-y)} \ket{p} = e^{-iqy} \ket{p+q}\0
\ee
and that
\be
\left(\int dq\, e^{-iqx} \ket{p+q}\right)\,*\,
\left(\int dq'\, e^{-iq'y} \ket{p+q'}\right)= \int dq\, e^{-iq(x+y)}
\ket{p+q}\label{x*y}
\ee
This clarifies the open string nature of the states (\ref{xp}). With
these results at hand one can now proceed to the explicit evaluation of
the LHS of (\ref{tensionbound}). The calculation in the open string
language now parallels exactly the one in the closed string language,
and will not be repeated here, see \cite{divecchia}. The final result is
the classic result (\ref{tensionbound}), provided
we make the identification
\be
\bra{\Xi_p} \Xi_p\rangle\sim T_p
\ee
This is expected if we keep in mind the relation with VSFT, since
the sliver (or the lump) are classical solutions having an energy that
reproduces the correct ratio of D--brane tensions.
One should also realize that this result can be the clue to understand why
the D--brane tension (as computed from the open string one loop computation
or, equivalently, from closed string exchange) is actually the same as the
energy density of the corresponding OSFT classical solution.

As a last remark, we would like to point out that our previous computation
is invariant under star rotations (since it is just
the computation of a bpz--norm) and so one can expect (in the complete theory
where the ghost sector is coupled consistently) this
to be a gauge invariant observable in OSFT.
At the moment, however, we are not able to give a physical
interpretation of what this gauge invariant object actually {\it is},
from a purely open string (field) theoretical point of view\footnote{In
perturbative open string theory the interpretation is in terms of one--loop
amplitude, but here the open string degrees of freedom appear
non--perturbatively, therefore the interpretation is expected to be different.
}.

\section{Discussion}

In this paper we have put forward a translation dictionary between
open and closed string theory in the framework of open string field theory.
We can summarize our proposal with the slogan: closed string modes are
star algebra projectors, where the star algebra is the one that appear in open
string field theory. Our starting point has been the identification
of the left and right sectors of the open string theory with the holomorphic
and antiholomorphic sectors of the closed string via a Bogoliubov transform.
The latter, in particular, maps the open string vacuum into the sliver string
field, which is identified with the closed string vacuum. We have shown that
zero momentum level--matched (off--shell) closed string states are associated
under our dictionary with star algebra projectors (or families thereof) in
the open string side. To associate a momentum to a given state we have to
shift to the lump vacuum and to tensor the previous states by a momentum
eigenstate which is itself a star algebra projector. So, altogether, we
can claim that according to our dictionary, off-shell closed string states
(i.e. momentum and level-matched closed string states) correspond to star
algebra projectors in the open string side.

We have presented one important outcome of our proposal, by showing that the
boundary state that represents a Dk--brane in the closed string language
is translated into the identity state in the open string side, which is
precisely the result one expects if our identification is correct. We
have tested this result by explicitly
showing how one can compute the closed string
exchange between two boundary states by using elementary star algebra
operations.

We also recall that the string states that in \cite{BMST2} were set in
correspondence with the 1/2 BPS LLM geometries, \cite{LLM}, turn out to be,
in the light of the present paper, infinite superpositions of closed
string states of the type (\ref{closedstate}) with $n=m$. This is another
element that fits the general scheme presented in this paper.

Of course this is only a beginning.
Many other tests have to be carried out and many problems have to be
clarified. To finish this paper we would like to make a list of the
impending issues.

Ghosts. We have to complete our dictionary with the inclusion of the ghost
sector. This is quite nontrivial because the analogy with the VSFT solutions
in this case is not very helpful. We recall that the VSFT equation of motion
for the ghost part is not a projector equation, while  one can expect our
ghost completion to be again related to projectors, in order to be
`bpz--dual' to the correct ghost number 3 boundary state.
This implies a nontrivial modification of the ghost
Neumann coefficients and ghost Fock space. We will deal with it in
a separate paper.

Role of the string midpoint. In open string field theory the string midpoint
plays a crucial role. In particular in VSFT, one could actually say that all
the physical observables are concentrated at that point (modulo singularities).
In the correspondence we have outlined in this paper the open string midpoint
does not seem to play any role: the reason can be traced back to the fact
that the Bogoliubov transformation is singular exactly at the midpoint,
so the latter has been swept away since the very beginning.
This seems to mark a basic difference between open and closed strings.

On--shell closed string states. In this paper we have considered
level and momentum--matched, but off--shell, closed string states.
The natural question is whether there is a simple
characterization of {\it on--shell} closed string states (i.e. states that
satisfy the full set of closed string Virasoro constraints) in terms
of open string modes. So far we have not been able
to find any appealing answer to this question.

Gauge freedom. As we have already remarked there is a large freedom in
choosing the basis $\xi_n$ or
$\xi'_n$, which we introduced in section 2 and 5, respectively. In fact this
freedom corresponds to an $O(\infty)$ group. Such large gauge freedom is
far from surprising in a string field theory context. We have already seen
that some relevant physical quantities (like the left/right levels or the
closed string exchange between two D--branes) are actually independent of
this choice.

The situation is more complicated when we come to other types of amplitudes.
For our dictionary may
allow us to calculate amplitudes between non--perturbative open string objects
and perturbative closed string modes: for instance, it may allow us to
compute the decay probability into the various closed string modes in the
process of a D--brane decay represented by a time--dependent
rolling tachyon--like solution, \cite{BMST1} (or, rather, by the corresponding
analytic solution \`a la Schnabl). This would allow us to identify the
tachyonic matter in a SFT context. Actually this has been the original
motivation of our research. The challenge in this
direction is precisely how to deal with the above large gauge freedom.
Since such gauge freedom is not present in open string theory, the gauge
freedom must be completely fixed. We have already started to do so by
choosing the basis as in (\ref{choice}), which was dictated by the physical
requirement that a boundary state in closed string theory should coincide
with the identity state in the open SFT side. However more comparisons like
this are needed between corresponding closed and open string objects
in order to fix the gauge freedom completely or, at least, to an acceptable
degree, which may allow us to do explicit calculations.

Closed string modes and analytic solutions of SFT. The basic objects
throughout our paper have been the star projectors. They are (for the matter
part) also solutions to the VSFT equations of motion. This property has not not
played any role in the above. However, as was
mentioned in the introduction, it may indicate that closed string modes
that correspond to star projectors might in fact correspond to full analytic
solutions to the SFT equation of motion. Verifying this may be crucial in
understanding open--closed string duality.

\vskip 1cm

\begin{center}
{\bf Acknowledgments}
\end{center}
N.B. and C.M. would like to thank G. Barnich, J. Evslin and  F. Ferrari
for discussions.\\
The research of L.B. is supported by the Italian MIUR under the
program ``Teoria dei Campi, Superstringhe e Gravit\`a''.\\
 N.B. is supported in part by a ``Pole d'Attraction
Interuniversitaire'' (Belgium), by IISN-Belgium, convention
4.4505.86, by Proyectos FONDECYT 1970151 and 7960001 (Chile) and
by the European Commission program
MRTN-CT-2004-005104, in which this author is associated to V.U. Brussels.\\
C.M. is supported in part by the Belgian Federal Science Policy Office
through the Interuniversity Attraction Pole P5/27, in part by the European
Commission FP6 RTN programme MRTN-CT-2004-005104 and in part by the
``FWO-Vlaanderen'' through project G.0428.06.\\

\subsection*{Appendix: diagonal representation of the $X$ and $X'$ matrices}

In this Appendix we collect some results, which are necessary in the text,
concerning the spectroscopy and diagonal representation of $X$ and $X'$
matrices.

The diagonalization of the $X$ matrix was carried out in \cite{RSZ4},
while the same analysis for $X'$ was accomplished in \cite{belov1} and
\cite{Feng}. Here, for later use, we summarize the results of
these references. The eigenvalues of $X=X^{11},X_+=X^{12},X_-=X^{21}$
and $T$ are given, respectively, by
\be
&& \mu^{rs}(k) = \frac{1-2\,\delta_{r,s}+e^{\frac{\pi k}2}\, \delta_{r+1,s}
+ e^{-\frac{\pi k}2}\, \delta_{r,s+1}}
{1+2\, {\rm cosh} {\frac{\pi k}2} }\label{muspec}\\
&&t(k) = - e^{-\frac {\pi |k|}2}\label{tspec}
\ee
where $-\infty<k<\infty$. The generating function for the eigenvectors is
\be
f^{(k)}(z) = \sum_{n=1}^\infty v_n^{(k)} \frac {z^n }{\sqrt{n}} =
\frac 1k (1-e^{-k\,{\rm arctan}\, z}) \label{fk}
\ee
The completeness and orthonormality equations for the  eigenfunctions
are as follows
\be
\sum_{n=1}^\infty v_n^{(k)}v_n^{(k')} = {\cal N}(k) \delta(k-k'),
\quad\quad {\cal N}(k) = \frac 2k \,{\rm sinh} \frac {\pi k}2,\quad\quad
 \int_{-\infty}^{\infty} dk\,\frac{v_n^{(k)} v_m^{(k)}}{{\cal N}(k)} =
\delta_{nm}\label{orth}
\ee
We define the normalized eigenvectors
\be
{\rm v}_n(k) = \frac {v_n^{(k)}}{\sqrt {\cal N}(k)}\0
\ee
and refer to ${\rm v}_n(k)$ as the {\it sliver basis}.

The spectrum of $X$ is continuous and lies in the interval $[-1/3,0)$. It is
doubly degenerate except at $-\frac 13$. The continuous spectrum of $X'$ lies
in the same interval, but $X'$ in addition has a discrete spectrum.
To describe it we follow \cite{belov1}. We consider the decomposition
\beq
X^{'rs} = \frac 13 (1+\a^{s-r} CU' + \a^{r-s} U'C )\label{decomp}
\eeq
where $\a=e^{\frac {2\pi i}3}$. It is convenient to express everything in
terms of $CU'$ eigenvalues and eigenvectors. The discrete
eigenvalues are denoted by $\xi$
and $\bar \xi$.
The matrix $CU'$ is hermitian, unitary and commutes with $U'C$.
Therefore $\xi$ and $\bar \xi$ lie on the unit circle and are
determined as follows, \cite{belov1}. Let
\be
\xi = - \frac{2-{\rm cosh} \,\eta- i \sqrt{3}\, {\rm sinh}\,\eta}
{1-2 {\rm cosh} \,\eta}
\label{xispec}
\ee
and
\be
F(\eta) = \psi\left(\frac 12 +\frac \eta{2 \pi i}\right) - \psi\left(\frac
12\right),
\quad\quad \psi(z)=\frac {d \log\Gamma(z)}{dz} \label{psi}
\ee
Then the eigenvalues $\xi$ and $\bar \xi$ are the solutions to
\be
\Re F(\eta)= \frac b4\label{eigeneq}
\ee
To each value
of $b$ there corresponds a couple of values of $\eta$ with opposite sign
(except for $b=0$ which implies $\eta=0$).

The eigenvectors $V_n^{(\xi)}$ are defined via the generating
function
\be
F^{(\xi)}(z) =\! && \sum_{n=1}^\infty V_n^{(\xi)} \frac {z^n}{\sqrt{n}}=
- \sqrt{\frac 2b} V_0^{(\xi)} \left[ \frac b4 +
\frac{\pi}{2\sqrt{3}}\, \frac{\xi-1}{\xi+1} + \log\, iz\,\right. \0\\
&&+ \left. e^{-2i(1+\frac{\eta}{\pi i}) {\rm arctan}\, z}
\Phi(e^{-4i\,{\rm arctan}\,z},1, \frac 12 +
\frac {\eta}{2 \pi i})\right]\label{genfund}
\ee
where $\Phi(x,1,y)= 1/y\,{}_2\!F_1(1,y;y+1;x)$, while
\beq
V_0^{(\xi)} = \left({\rm sinh}\,\eta \frac {\d}{\d \eta}
\left[\log \Re F(\eta)\right]\right)^{-\frac 12}\label{V0xi}
\eeq

As for the continuous spectrum, it is spanned by the variable $k$,
$-\infty<k<\infty$.
The eigenvalues of $CU'$ are given by
\be
\nu(k)= -\frac {2 +{\rm cosh}\, \frac{\pi k}{2} + i \sqrt{3}\, {\rm sinh}
\,\frac{\pi k}{2} }{1+ 2\,{\rm cosh }\, \frac{\pi k}{2} }\0
\ee
The generating function for the eigenvectors is
\be
&&F_c^{(k)}(z) = \sum_{n=1}^{\infty} V_n^{(k)} \frac {z^n}{\sqrt{n}} =
V_0^{(k)}\sqrt{\frac 2b} \left[ -\frac b4 -\left(\Re F_c(k)-\frac b4\right)
e^{-k\,{\rm arctan}\,z}-\log\, iz\right.\label{genfunc}\\
&&-\left.\left(\frac{\pi}{2\sqrt{3}}\,\frac{\nu(k)-1}{\nu(k)+1}+
\frac {2i}k\right)
+2i\,f^{(k)}(z) -\Phi(e^{-4i\,{\rm arctan}\,z},1,1  +
\frac {k}{4 i})\,e^{-4i\,{\rm arctan}\,z}\,e^{-k\,{\rm arctan}\,z}
\right]\0
\ee
where
\be
F_c(k) = \psi(1 +\frac k{4 \pi i}) - \psi(\frac 12)\0,
\ee
while
\be
V_0^{(k)} = \sqrt{\frac b{2{\cal N}(k)}}
\left[4 +k^2\left(\Re F_c(k)-\frac b4\right)^2\right]^{-\frac 12}
\label{V0k}
\ee

The continuous eigenvalues of $X',X_-',X_-'$ and $T'$ (for the conventional
lump) are given by same
formulas as for the $X,X_+,X_-$ and $T$ case, eqs(\ref{muspec},\ref{tspec}).
As for the discrete eigenvalues, they are given by the formulas
\be
&& \mu^{rs}_\xi =\frac {1-2\,\delta_{r,s}-e^\eta\,
\delta_{r+1,s}-e^{-\eta}\,\delta_{r,s+1}}{1-2\,{\rm cosh}\,\eta} \0\\
&&t_\xi= e^{-|\eta|}\label{xtspec}
\ee

The eigenvectors corresponding to the continuous spectrum are
$V_N(k)$  ($-\infty<k<\infty$), while the eigenvectors of the discrete
spectrum are denoted by  $V_N(\eta)$ and  $V_N(-\eta)$ .
They form a complete basis.
They will be normalized so that the completeness relation takes the form
\be
\int^{\infty}_{-\infty} dk\, V_N(k)V_M(k) + V_N(\eta)V_M(\eta)
+V_N(-\eta)V_M(-\eta) = \delta_{NM}\label{completelump}
\ee
We refer to $V_N$ as the {\it lump basis}.

One important difference between the ${\rm v}_n$ and $V_N$ basis is determined
by the twist transformation properties (in vector notation, ${\rm v}=
\{{\rm v}_n\}$, etc.)
\be
C{\rm v}(k) = -{\rm v}(-k), \quad\quad {\rm while} \quad\quad CV(k)=
V(-k)\label{difference}
\ee
We have also $CV(\eta)=V(-\eta)$.


\end{document}